\documentclass[12pt]{iopart}
\usepackage{graphics}

\begin{document}
\article[Thermal and Curvature Effects to Spontaneous Symmetry Breaking
 in $\phi^4$ Theory]{}{Thermal and Curvature Effects to Spontaneous Symmetry Breaking
 in $\phi^4$ Theory}
\author{T Hattori$^1$, M Hayashi$^1$, T Inagaki$^2$ and Y Kitadono$^1$}
\address{$E$ Department of Physics, Hiroshima University, 
Higashi-Hiroshima, Hiroshima, 739-8521, Japan}
\address{$E$ Information Media Center, Hiroshima University, 
Higashi-Hiroshima, Hiroshima, 739-8521, Japan}
\eads{\mailto{tkc@hirohisma-u.ac.jp},\mailto{hayashi@theo.phys.sci.hiroshima-u.ac.jp}, 
\mailto{inagaki@hiroshima-u.ac.jp}, \mailto{kitadono@theo.phys.sci.hiroshima-u.ac.jp}}
\begin{abstract}
We study the thermal and curvature effect to spontaneous
symmetry breaking in $\phi^4$ theory. The effective potential
is evaluated in  $D$-dimensional static universe with positive
curvature $R\otimes S^{D-1}$ or negative curvature $R\otimes H^{D-1}$. 
It is shown that temperature and positive curvature suppress the symmetry breaking, 
while negative curvature enhances it. To consider the back-reaction
we numerically solve the gap equation and the Einstein equation 
simultaneously. The solution gives the relationship between the 
temperature and the scale factor.
\end{abstract}

\section{Introduction}
Spontaneous symmetry breaking is one of the important concept in elementary 
particle physics. In early universe, especially the grand unification era 
($\sim 10^{-34}$s), the strong curvature and high temperature state is realized. 
We can not neglect the thermal and curvature effects to the cosmic evolution. 
We need to treat the matter as quantized fields. However, we assume that  
it is adequate to consider the gravity as a classical field in GUT era. 
The thermal and curvature effects to spontaneous symmetry breaking 
has been studied by many authors \cite{Chen:1985wb, Hu:1986jh, Roy:1989gj, DC}. 
The curvature effect to the symmetry breaking comes from a scale factor dependence of 
the covariant derivative and a coupling between the scalar field and the gravitational field.
One-loop and two-loop corrections to scalar field theories in the
linear curvature approximation were found in Refs. \cite{BO1}
and \cite{BO2}, respectively. 
The relationship between the scale factor and the temperature for a free scalar field was 
calculated by solving the back reaction problem in Refs. \cite{AS,BA}. 

In the present article, we extend the work in Refs. \cite{AS,BA} to the $\phi^4$ 
theory and investigate the structure of early universe. 
For this purpose we study the thermal and curvature effects to the spontaneous 
symmetry breaking in a D-dimensional static universe. 
To obtain finite results we confine ourselves to the spacetime dimensions 
$2\leq D<5$. As a simple example, we consider the scalar $\phi ^4$ 
theory where the $Z_2$ symmetry is spontaneously broken. The explicit expression 
of the 1-loop effective potential of $\phi ^4$ theory is derived on a positive 
curvature spacetime $R \otimes S^{D-1}$ or a negative curvature spacetime 
$R \otimes H^{D-1}$ \cite{IIM1}. We find the phase structure of the model by observing the 
minimum of the effective potential with varying temperature and scale factor in our 
previous work \cite{Pot}. 

The expectation value of the stress-tensor, the right-hand side of the Einstein equation, 
is also calculated. We numerically solve the gap equation and the Einstein equation 
simultaneously on $R \otimes S^{D-1}$ and $R \otimes H^{D-1}$ at finite temperature. 
The solution gives the relationship between the temperature and the scale factor of the 
universe.

\section{Spontaneous symmetry breaking in $R \otimes S^{D-1}$ and $R \otimes H^{D-1}$ at finite 
temperature}
In this section, we briefly review the analysis by the effective potential in curved 
spacetime at finite temperature and observe the property of phase transition with varying 
temperature and scale factor. 

We introduce the constant curvature space $R \otimes S^{D-1}$ and $R \otimes H^{D-1}$ as 
Euclidean analog of the static universe. The manifold $R \otimes S^{D-1}$ is represented 
by the metric 
\begin{equation}
	ds^2 = dr^2+a^2(d\theta^2+\sin^2\theta d\Omega_{D-2}). 
\end{equation}
where $d\Omega_{D-2}$ is the metric on a unit sphere $S^{D-2}$ and $a$ is the scale factor. 
Similarly, the manifold $R \otimes H^{D-1}$ is defined by the metric 
\begin{equation}
	ds^2 = dr^2+a^2(d\theta^2+\sinh^2\theta d\Omega_{D-2}). 
\end{equation}
The Lagrangian of the scalar $\phi^4$ theory is written as 
\begin{equation}
  {\cal L}(\phi ) = - \frac{1}{2}\phi (\opensquare + \xi_0 R )\phi
  +\frac{\mu_0^2}{2}\phi^2 
  -\frac{\lambda_0}{4!}\phi^4, 
\end{equation}
where $\phi$ is a real scalar field, $i\mu_0$ corresponds to the bare mass of the scalar 
field, $\lambda_0$ the bare coupling constant for the scalar self-interaction, $\xi_0$ 
the bare coupling constant between the scalar field and the gravitational field. 

According to Ref. \cite{Pot}, in Euclidean spacetime the effective potential $V(\phi)$ can 
be written as 
\begin{equation}
	V(\phi ) = \frac{\xi_0}{2} R\phi^2 + \frac{\lambda_0}{4!}\phi^4+\frac{\hbar \lambda_0}{4}
	\int_0^{\phi^2} dm^2 G(x,x;m) + O(\hbar^{3/2}). 
\label{pot:FT}
\end{equation}
In this article we neglect $O(\hbar^{3/2})$ term.
The two-point function $G(x,y)$ of the real scalar field at finite temperature 
on the manifolds $R\otimes S^{D-1}$ and on the manifolds $R \otimes H^{D-1}$ are given 
by Ref. \cite{IIM1}.

The effective potential $V(\phi)$ obtained in the previous section is divergent 
at D=2 or D=4. We introduce the renormalization procedure by imposing the 
renormalization conditions 
\begin{equation}
  \left. \frac{\partial^2 V_0}{\partial \phi^2} \right|_{\phi=0}
  \equiv -\mu_r^2-\xi_r R , \hspace{2ex}
  \left. \frac{\partial^4 V_0}{\partial \phi^4} \right|_{\phi=M}
  \equiv \lambda_r ,
\end{equation}
where $M$ is the renormalization scale and $V_0 (\phi)$ is the effective 
potential in flat spacetime at $T=0$. The renormalized effective potential is 
given by replacing the bare parameters $\mu_0$ and $\lambda_0$ with $\mu_r$ and $\lambda_r$ 
in the tree level terms in $V(\phi)$. 

The behavior of the effective potential is illustrated in our previous paper \cite{Pot}.
The curvature effect suppresses the symmetry breaking in a positive 
curvature space $R\otimes S^{D-1}$, while it enhances the symmetry breaking in a negative
curvature space $R\otimes H^{D-1}$.

\section{Back reaction on $R \otimes S^{D-1}$ and $R \otimes H^{D-1}$ at finite temperature}

In a quantum field theory the Einstein's field equation is given by 
\begin{equation}
	R_{\mu\nu} -\frac{1}{2}Rg_{\mu\nu} = 8\pi \langle T_{\mu\nu} \rangle , 
\label{EinsteinEq}
\end{equation}
where $\langle T_{\mu\nu}\rangle$ is the expectation value
of the stress tensor in the ground state which is determined
by observing the minimum of the effective potential. 
Units have been chosen so that $G = c = \hbar = k = 1$ in this section. 

To include the back reaction we solve the gap equation 
\begin{equation}
  \left. \frac{\delta V(\phi)}{\delta \phi} \right|_{\phi \rightarrow \langle \phi \rangle} = 0.
\label{Gap:E}
\end{equation}
and the $44$ component of \Eref{EinsteinEq} simultaneously. Since $R_{44}=0$
in $R\otimes S^{D-1}$ and $R\otimes H^{D-1}$, the $44$ component
of the Einstein equation reduces to
\begin{equation}
	-\frac{1}{2}R = 8\pi \langle T_{44} \rangle . 
\label{E44}
\end{equation}
By the variation over the metric, we obtain $\langle T_{44}\rangle$
on $R\otimes S^{D-1}$
\begin{eqnarray}
	\langle T_{44}\rangle &=& \frac{1}{2}(-\mu_0^2 + \xi_0 R)\langle \phi \rangle^2 
		+ \frac{\lambda_0}{4!}\langle \phi \rangle^2 
		- \frac{1}{(4\pi )^{\frac{D-1}{2}}}\Gamma \left( \frac{3-D}{2} \right)Ta^{3-D} \nonumber \\
	&& \times \int_{0}^{\infty} \sum^{n=\infty}_{n=-\infty}\omega_n^2
		\frac{\Gamma\left( \frac{D-2}{2}+i\alpha_S \right)
		\Gamma\left( \frac{D-2}{2}-i\alpha_S \right)}{\Gamma\left( \frac{1}{2}+i\alpha_S \right)
		\Gamma\left( \frac{1}{2}-i\alpha_S \right)}, 
\label{ST44}
\end{eqnarray}
and on $R \otimes H^{D-1}$, 
\begin{eqnarray}
	\langle T_{44}\rangle &=& \frac{1}{2}(-\mu_0^2 + \xi_0 R)\langle \phi \rangle^2 
		+ \frac{\lambda_0}{4!}\langle \phi \rangle^2 
		- \frac{1}{(4\pi )^{\frac{D-1}{2}}}\Gamma \left( \frac{3-D}{2} \right)Ta^{3-D} \nonumber \\
	&& \times \int_{0}^{\infty} \sum^{n=\infty}_{n=-\infty}\omega_n^2
		\frac{\Gamma\left( \frac{D-2}{2}+\alpha_H \right)}
		{\Gamma\left( \frac{4-D}{2}+\alpha_H \right)}. 
\label{HT44}
\end{eqnarray}

\begin{figure*}
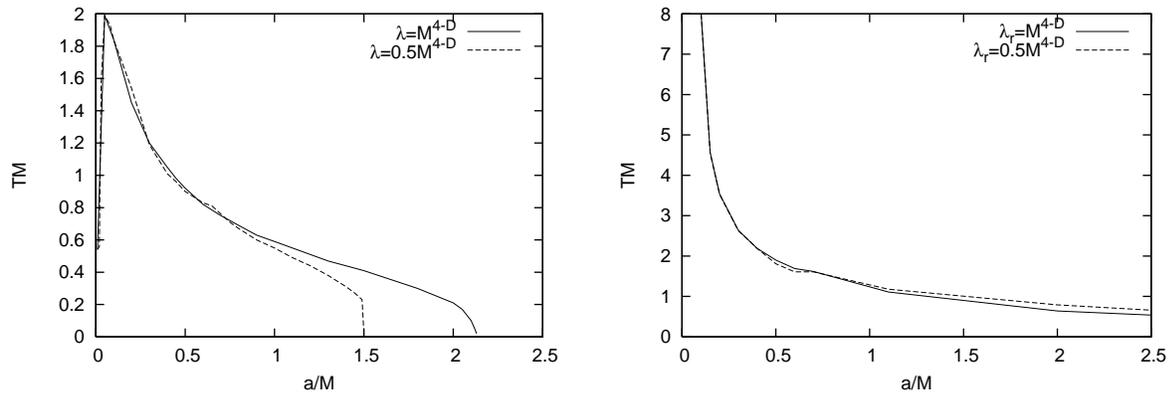
 
\begin{minipage}{8cm}
\includegraphics{figure2a.eps}

(a) {\scriptsize $R \otimes S^{D-1}$, $\mu^2=0.1M^2$, $\lambda=M^{4-D}, 0.5M^{4-D}$}
\end{minipage}
\begin{minipage}{8cm}
\includegraphics{figure2b.eps}

(b) {\scriptsize $R \otimes H^{D-1}$,$\mu^2=0.1M^2$, $\lambda=M^{4-D}, 0.5M^{4-D}$}
\end{minipage}
\caption{\label{aT} The $T-a$ relationship for a conformally 
coupled scalar field $\xi=(D-2)/(4D-4)$.}
\end{figure*}

We numerically solve the gap equation \eref{Gap:E} and the $44$ component 
of the Einstein equation \eref{E44}. The solution gives the relationship between temperature 
and radius of the universe. We show for the self-interacting fields, $\lambda_r=M^{4-D}$ and 
$\lambda_r=0.5M^{4-D}$ in Fig.\ref{aT}. The scalar self-interaction increases the scale factor $a$ on 
$R \otimes S^{D-1}$, while it decreases $a$ on $R \otimes H^{D-1}$. 

\section{Concluding remarks}
We have investigated the thermal and curvature effect to spontaneous symmetry 
breaking at finite temperature and curvature in arbitrary dimensions ($2\leq D<5$). 

In a positive curvature space $R \otimes S^{D-1}$ the broken symmetry is restored as 
the curvature increases. For a negative curvature space $R \otimes H^{D-1}$ we see 
that the symmetry breaking is enhanced by the curvature effect. 
Solving the gap equation and the Einstein equation, we show the relationship between 
the temperature and the scale factor. By the scalar self-interaction, the scale factor 
is increased on $R \otimes S^{D-1}$, whereas it is decreased on $R \otimes H^{D-1}$. 

It is important to extend our analysis to the time dependent back ground. Thus it is 
important to consider the model in the state out of equilibrium. 
We remained it for future works.

\end{document}